\documentclass{PoS}

\title{Moments of generalized parton distributions and quark angular momentum of the nucleon}

\ShortTitle{Moments of GPD and quark angular momentum of the nucleon}

\author{Dirk Br\"ommel$^{ab}$, 
Meinulf G\"ockeler$^b$, Philipp H\"agler$^c$, Roger Horsley$^d$,
    Yoshifumi~Nakamura$^e$, \speaker{Munehisa Ohtani}$^b$,
Dirk Pleiter$^e$, Paul E.L. Rakow$^f$, Andreas~Sch\"afer$^b$,
  Gerrit Schierholz$^{ae}$, Wolfram Schroers$^e$, 
Hinnerk St\"uben$^g$, and James M.~Zanotti$^d$ 
 \\
\llap{$^a$} Deutsches Elektronen-Synchrotoron DESY, 22603 Hamburg, Germany \\
\llap{$^b$} Institut f\"ur Theoretische Physik, Universit\"at Regensburg, 93040
Regensburg, Germany \\
\llap{$^c$} Institut f\"ur Theoretische Physik T39, Physik-Department der 
TU M\"unchen, James-Franck-Stra{\ss}e, 85747 Garching, Germany\\
\llap{$^d$} School of Physics, University of Edinburgh, Edinburgh EH9 3JZ, UK\\
\llap{$^e$} John von Neuman-Institut f\"ur Computing NIC/DESY, 15738 Zeuthen, Germany\\
\llap{$^f$} Theoretical Physics Division, Department of Mathematical Sciences,
 University of Liverpool, Liverpool L69 3BX,UK \\
\llap{$^g$} Konrad-Zuse-Zentrum f\"ur Informationstechnik Berlin,
14195 Berlin, Germany

\\        E-mail: \email{munehisa.ohtani@physik.uni-regensburg.de}}
\author{QCDSF-UKQCD Collaboration}

\abstract{The internal structure of hadrons is important for a variety of  
topics, including the hadron form  factors, proton spin and spin asymmetry 
in polarized proton  
scattering. For a systematic study generalized parton distributions  
(GPDs) encode important information on hadron structure in the entire  
impact parameter space.
We report on a computation of nucleon GPDs based on simulations with two  
dynamical non-perturbatively improved Wilson quarks with pion masses  
down to 350MeV. We present results for the total angular momentum of  
quarks with chiral extrapolation based on covariant baryon
chiral perturbation theory.}

\FullConference{The XXV International Symposium on Lattice Field Theory\\
                 July 30 - August 4 2007\\
                 Regensburg, Germany}

\begin{document}

\section{Introduction}
 The internal structure of nucleons has attracted much attention in 
the contexts of
the nucleon form factors, proton spin 
and spin/charge asymmetry in deeply virtual Compton scattering and so on. 
For a systematic study
of the nucleon internal structure, generalized parton distributions (GPDs) are
introduced through the off-forward matrix elements of quark-bilinear operators:
\begin{equation}
\int\!\!\frac{d\eta}{4\pi}e^{i\eta x}\langle P'| 
\bar{q}({\textstyle -\frac{\eta n}{2}})
\gamma^\mu 
{\cal U} 
q({\textstyle\frac{\eta n}{2}}) |P\rangle
= \bar{N}(P')\!\!\left(\gamma^\mu 
H(x,\xi,t)
+ i{\textstyle \frac{\sigma^{\mu\nu}{\Delta_\nu}}{2M}}
E(x,\xi,t)
 \right)\!\! N(P),
\end{equation}
with a light cone vector $n$ and the momentum transfer $\Delta=P'-P$ 
as functions of the quark momentum fraction $x$, the skewedness 
$\xi=-n\cdot\Delta/2$ and the virtuality $t=\Delta^2$.
The axial counterparts are denoted by $\tilde{H}$ and $\tilde{E}$.
Since the GPD is defined with the finite momentum transfer in contrast to
the conventional parton distribution functions, 
partons bring us the informations on hadron structure in the transverse space.

In this contribution,
we report on the first moments of GPD, so called generalized form factors, 
for nucleon, as a function of the virtuality calculated on the lattice
with unquenched configurations of QCDSF/UKQCD collaboration.

 In the forward limit these generalized form factors provide the total 
angular momentum of quark in the nucleon through Ji's sum rule \cite{Js},
\begin{equation}
  J^q=\frac{1}{2}\int_{-1}^1 dx x (H(x,\xi,0)+E(x,\xi,0))
 \equiv\frac{1}{2}(A_{20}(t=0)+B_{20}(t=0)).
\end{equation}
Combined with the quark spin contributions to the nucleon
obtained as the forward value of the axial form factor,
\begin{equation}
  s^q=\frac{1}{2}\int_{-1}^1 dx  \tilde{H}(x,\xi,0)
\equiv\frac{1}{2}\tilde{A}_{10}(t=0),
\end{equation}
we compute the orbital angular momentum of quarks as $L^q=J^q-s^q$. 
Using the results of chiral perturbation theory ($\chi$PT)
for chiral extrapolation to the physical point,
we discuss the angular momentum carried by quark in the nucleon.

\section{Generalized form factors on the lattice}

The Mellin moments of the GPDs are known to be expressed by
polynomials in terms of $\xi$ \cite{XJ},
\begin{equation}
\int_{-1}^1 dx x^{n-1} \left[\begin{array}{c}{ 
{H}(x,\xi,t)} \\ {E(x,\xi,t)} \end{array}\right]
= \sum_{k=0}^{[(n-1)/2]}
(2\xi)^{2k} \left[\begin{array}{c}
{ A_{n,2k}(t)} \\{ B_{n,2k}(t)}\end{array}\right] 
\pm \delta_{n,\rm even}(2\xi)^{n}{ C_{n}(t)}.
\end{equation}
The generalized form factors $A_{n,2k},B_{n,2k}$ and 
$C_{n}$ are defined from the coefficients of this expansion.
Since the integration by $x$ makes the quark operator local,
the $(n-1)$-th moments can be calculated \cite{QL} on the lattice
through the matrix element of 
 $\langle P'|\bar{q}\gamma^{\{\mu_1}D^{\mu_2}\cdots D^{\mu_n\}}q|P\rangle$
by taking a ratio of the three- and two-point functions.

To estimate these correlation functions, 400 to 2200 configurations 
are used for each $\beta, \kappa$
with two flavor Wilson fermion with the clover improvement.
Simulations are performed with various set of parameters $\beta$ and $\kappa$
corresponding to the lattice spacing less than 0.09fm and pion mass
covering from order of 1GeV down to 350MeV with a reference scale
$r_0=0.467$fm.
Nonperturbative renormalizations are incorporated to convert the lattice
results into the values in the $\overline{\rm MS}$ scheme at a scale of 
$\mu^2=4$GeV$^2$.
The ${\cal O}(a)$ improvement of the quark energy-momentum tensor 
are carried out through the boosted perturbation theory following 
ref.\cite{bpt}
and the tadpole improved version is used for the axial current
following ref.\cite{ga}.
 We note that the contributions from disconnected diagrams 
are not included in the present lattice results.

\section{Lattice simulation results and chiral extrapolation}

We focus on the generalized form factors $A_{20}$ and $B_{20}$
as well as the axial form factor $\tilde{A}_{10}$ to evaluate the 
quark angular momentum in the nucleon.

\begin{figure}[t!]
\centering
{\includegraphics[scale=.33,angle=270]{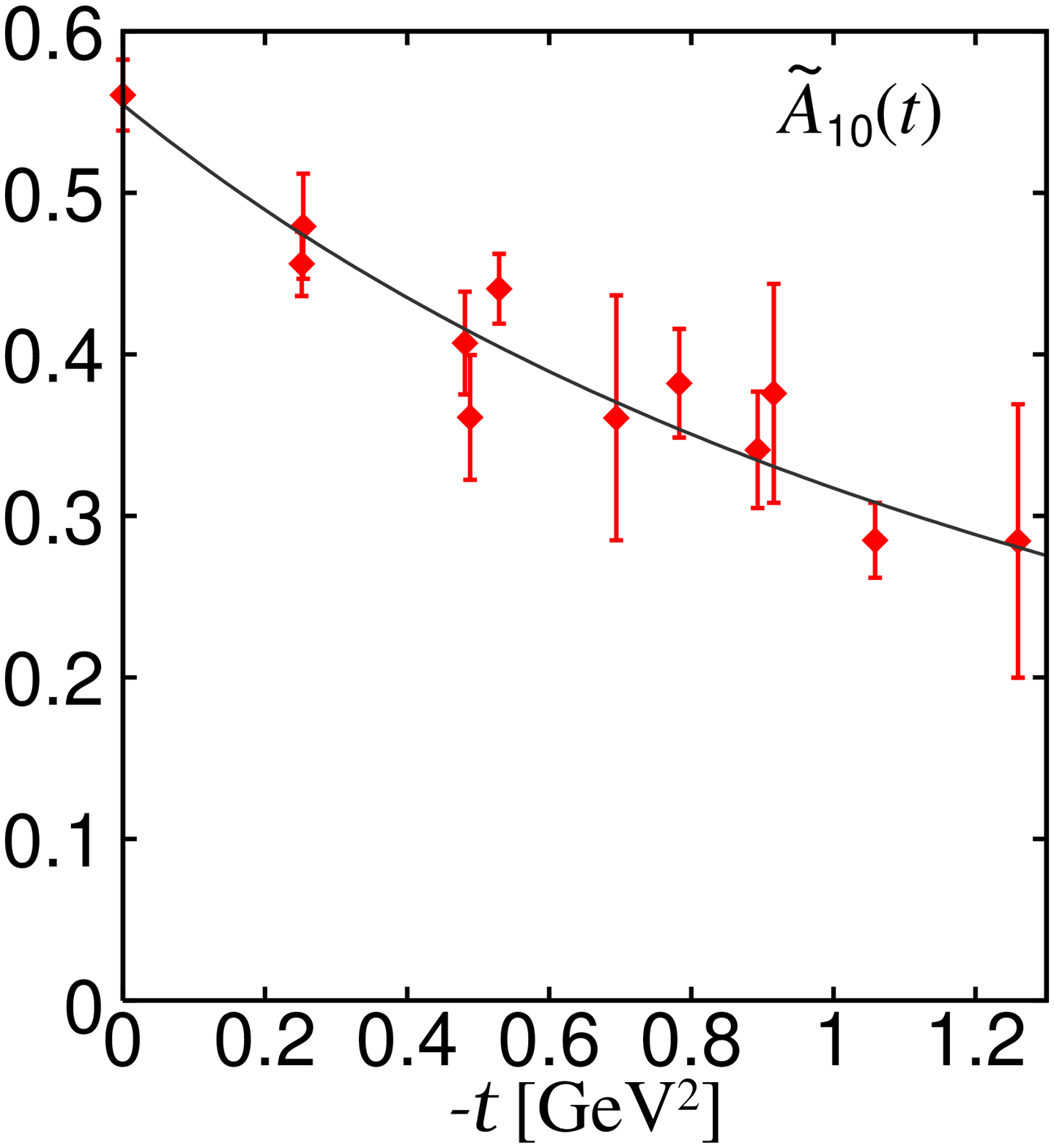} \hspace*{1.6em}
\includegraphics[scale=.33,angle=270]{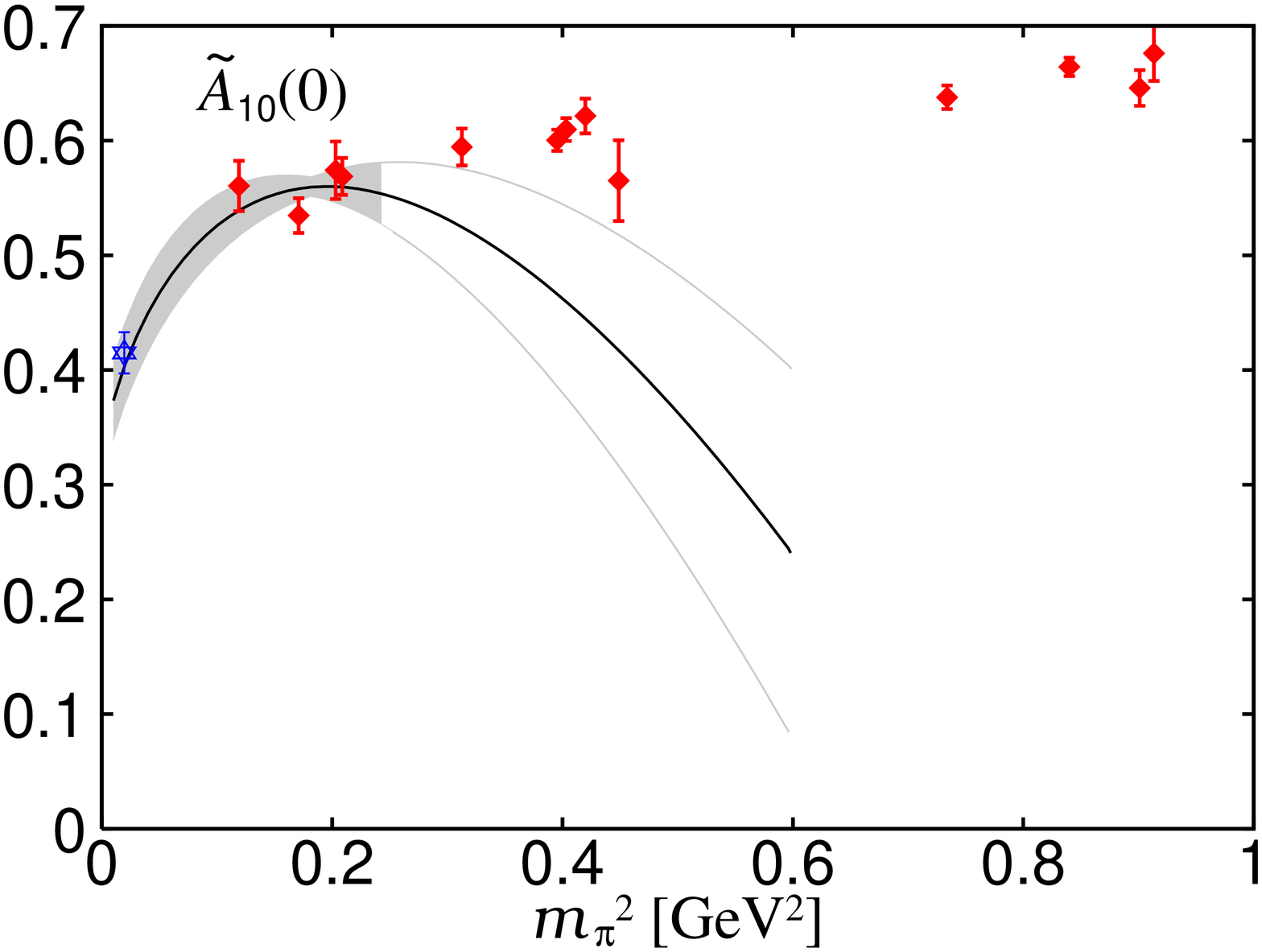}}
  \caption{ The isoscalar axial form factor $\tilde A_{10}(t)$ 
for $\beta=5.29$, $\kappa=0.13632$
with dipole fit (left) and the forward values with $\chi$PT fit (right).
The open star in the right panel represents the latest experimental
value of HERMES.}
  \label{fig:DS}
\end{figure}

Typical $t$ dependence of the axial form factor in the isoscalar channel
are shown in Fig.\ref{fig:DS}. 
The obtained lattice data agrees well with
a fitting by the dipole form, $\tilde{A}_{10}(0)/(1-t/m^2)^2$.
The forward values obtained by setting $t=0$ present a smooth pion-mass
dependence as shown in the right panel of Fig.\ref{fig:DS}.

Here we use an expression derived in a heavy baryon $\chi$PT \cite{DMS},
\begin{equation}
 \tilde{A}_{n,k}^{\rm u+d}(0)={ \alpha_{n,k}}\left[
1-\frac{3m_\pi^2g_A^2}{16\pi^2F_\pi^2}\left(
\ln\frac{m_\pi^2}{\lambda^2}+1\right)
\right]+{\beta_{n,k}}m_\pi^2+{\cal O}(m_\pi^3),
\end{equation}
for the chiral extrapolation with fitting parameters $\alpha_{10}$ and 
$\beta_{10}$ at a scale of $\lambda=1$GeV.
As the heavy baryon formalism is valid only for the small pion mass,
we restrict the data points at pion masses less than 
500MeV.
Then it turns out that the chiral log term gives a strong $m_\pi$
dependence for small $m_\pi$ region and the extrapolated
value is eventually comparable with the latest experimental value
of deep-inelastic scattering reported by HERMES \cite{HER}.
With this extrapolation, we obtain the quark spin contribution 
in the nucleon as 
$\tilde{A}_{10}^{\rm u+d}(0)\equiv 2s^{\rm u+d} = 0.402\pm 0.024$
at the physical pion mass.

\begin{figure}[t!]
\centering
{\includegraphics[width=.35\textwidth,angle=270]{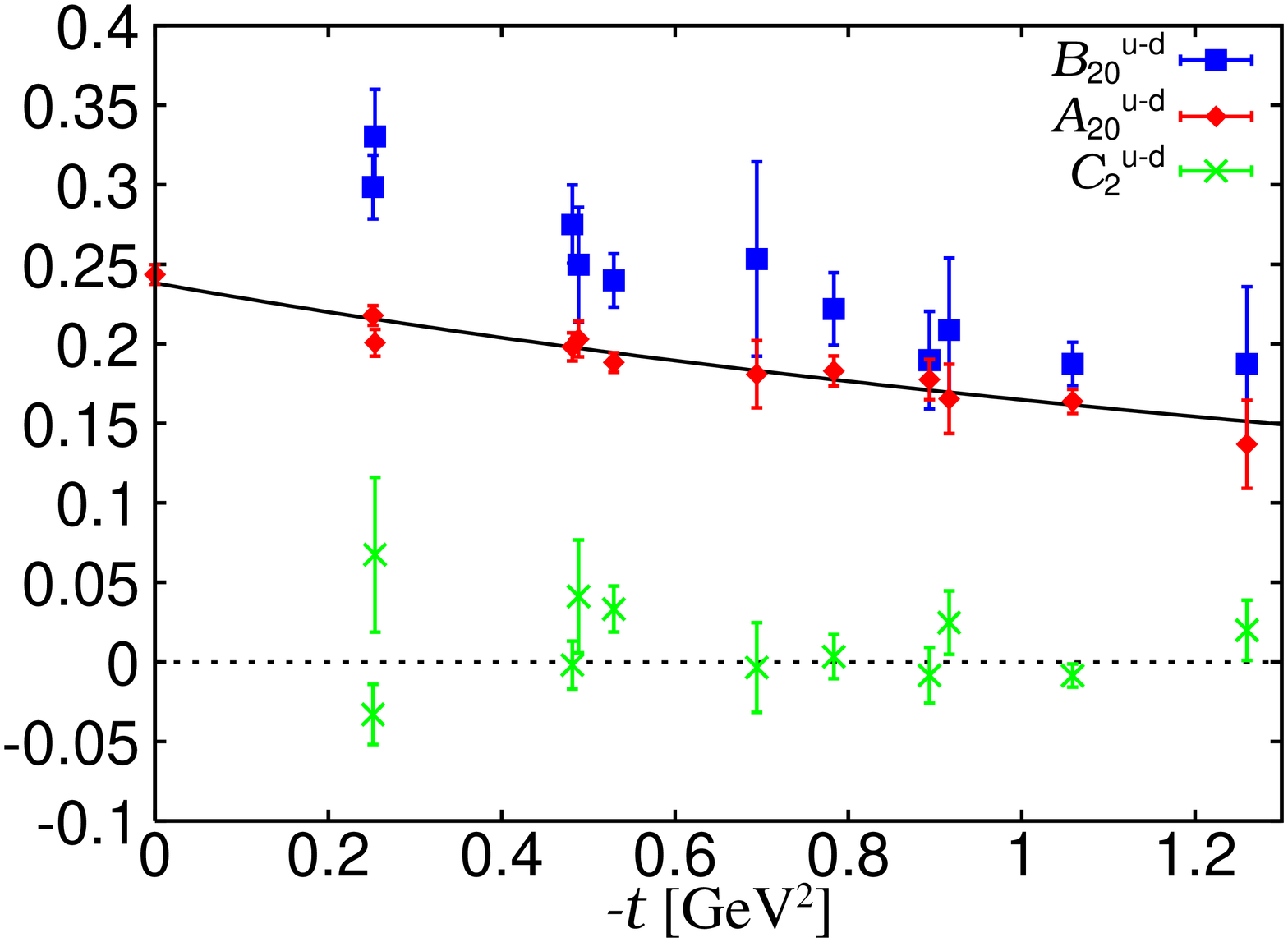}\hspace*{1.6em}
\includegraphics[width=.35\textwidth,angle=270]{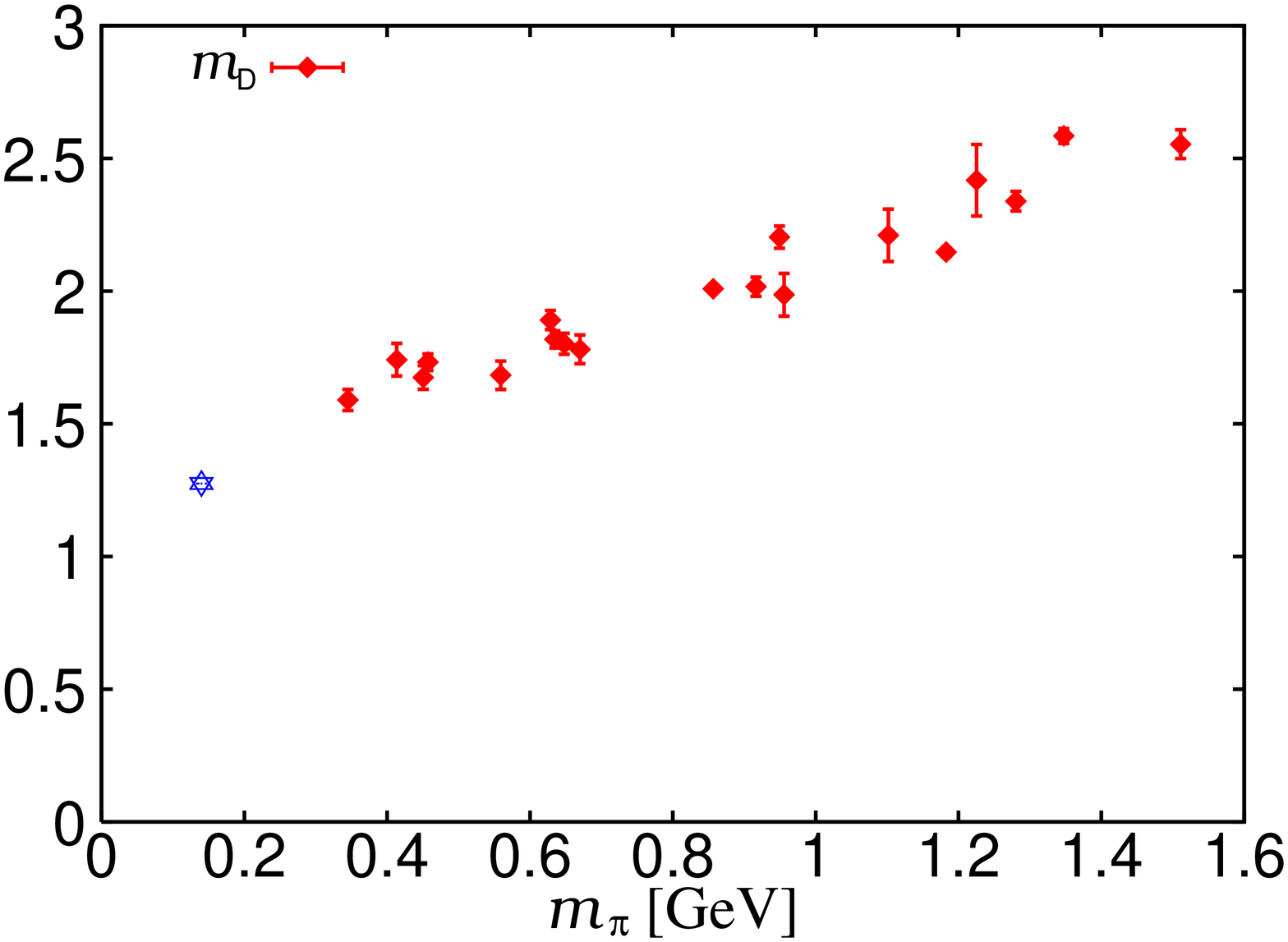}}
  \caption{ Generalized form factors in the isovector channel
for $\beta=5.29$, $\kappa=0.13632$ with dipole fit for $A_{20}$
and the dipole mass of $A_{20}$.
The open star represents the experimental value of $f_2$ tensor meson mass.}
  \label{fig:GFF}
\end{figure}

Next we show a typical $t$ dependence of generalized form factors
$A_{20},B_{20}$ and $C_{2}$ in the isovector channel in Fig.\ref{fig:GFF}.
Up to 1GeV$^2$, the lattice results of $A_{20}$ 
agree with the dipole fit $A_{20}(0)/(1-t/m_{\rm D}^2)^2$.
The dipole mass $m_{\rm D}$ of $A_{20}$ shows a 
smooth $m_\pi$ dependence shown in the right panel of Fig.\ref{fig:GFF}.
We see that the dipole mass seems to extrapolate
to the observed mass of tensor meson $f_2$ at the physical point.
This fact contrasts with a mass scale of the electromagnetic form factors
comparable with the vector meson mass \cite{WS}.

However, the empirical dipole fit
for the generalized form factors has no solid justification from
a theoretical point of view. 
Therefore we count on the covariant baryon chiral perturbation \cite{DGH}
to extract the forward values of $B_{20}$ as well as the chiral
extrapolation of $A_{20}$ and $B_{20}$.

\begin{figure}[b!]
\centering
{\includegraphics[width=.35\textwidth,angle=270]{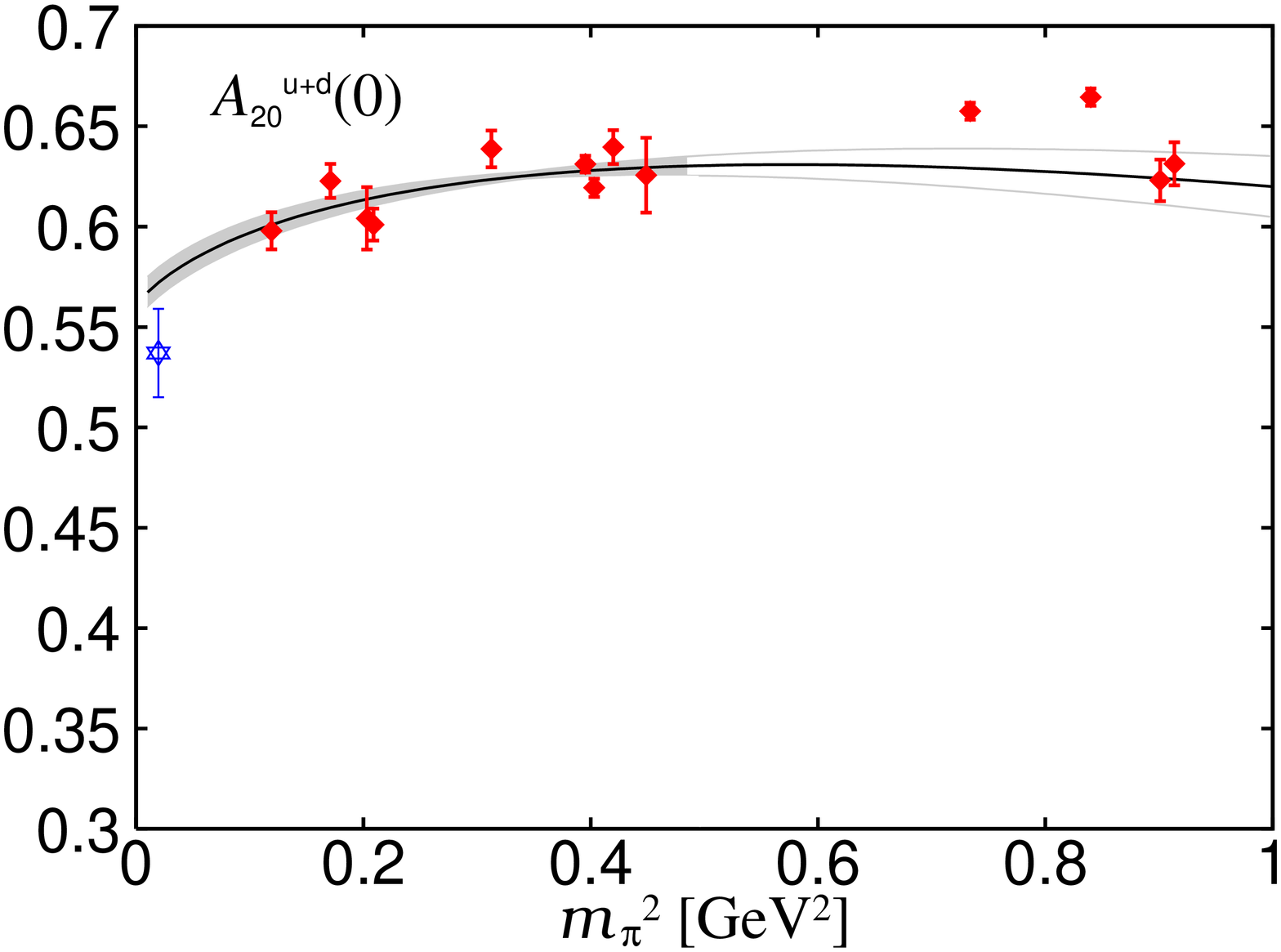} \hspace*{1.6em}
\includegraphics[width=.35\textwidth,angle=270]{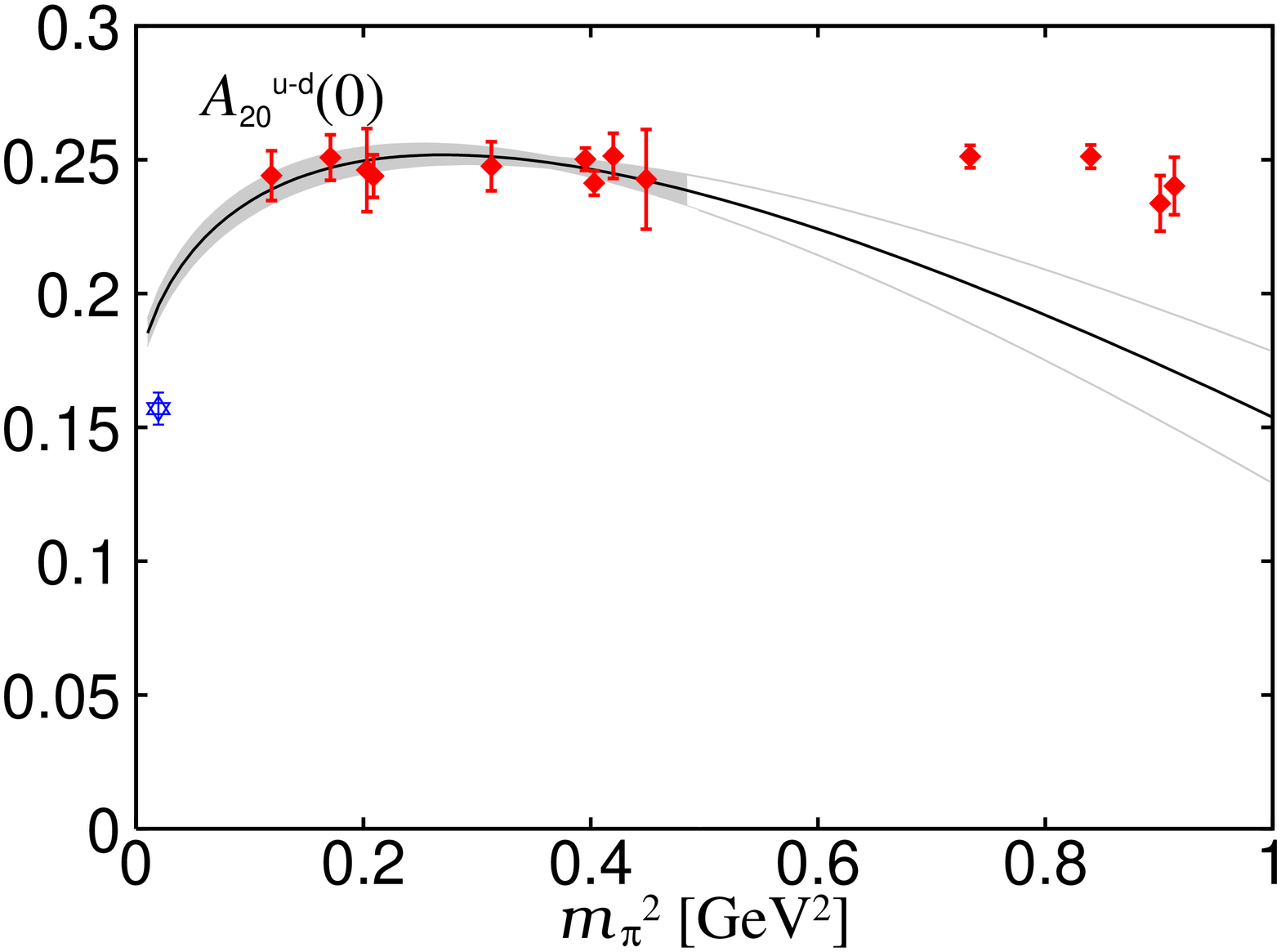}}
  \caption{ The forward values of $A_{20}$ in the 
isoscalar (left) and isovector (right) channel with $\chi$PT fits. 
The open stars represent the phenomenological values from CTEQ6.}
  \label{fig:A20}
\end{figure}

 The forward values of $A_{20}$ are identical to 
the quark momentum fraction $\langle x\rangle^q$ and
are shown as a function of $m_\pi^2$ in Fig.\ref{fig:A20}.
Both in the isoscalar and isovector channel, the lattice results show
a moderate pion mass dependence.
These values are extrapolated to the physical point by 
the following eqs. derived in the baryon $\chi$PT, 
\begin{eqnarray}
A_{2,0}^{\rm u+d}(0)&=&{ a_{20}^{\rm s}}+{ c_9}
\frac{4m_\pi^2}{M_0^2} 
-{ a_{20}^{\rm s}}
\frac{3g_A^2m_\pi^2}{16\pi^2F_\pi^2}\left[
\frac{m_\pi^2}{M_0^2}+\frac{m_\pi^2}{M_0^2}\left(
2-\frac{m_\pi^2}{M_0^2}\right)\ln\frac{m_\pi}{M_0}
\right.\nonumber \\ & & \hspace*{.3em} \left. 
+\frac{m_\pi}{\sqrt{4M_0^2-m_\pi^2}}\left(
2-4\frac{m_\pi^2}{M_0^2}+\frac{m_\pi^4}{M_0^4}
\right)\arccos\frac{m_\pi}{2M_0}
\right] +{\cal O}(p^3),
\end{eqnarray}
for the isoscalar channel and
\begin{eqnarray}
A_{2,0}^{\rm u-d}(0)&=&{ a_{20}^{\rm v}}+{ c_8}
\frac{4m_\pi^2}{M_0^2}
+{ a_{20}^{\rm v}}
\frac{g_A^2m_\pi^2}{16\pi^2F_\pi^2}\left[
-\left(3+\frac{1}{g_A^2}\right)\ln\frac{m_\pi^2}{\lambda^2}+
\frac{m_\pi^2}{M_0^2}-2+\frac{m_\pi^2}{M_0^2}\left(
6-\frac{m_\pi^2}{M_0^2}\right)\ln\frac{m_\pi}{M_0}
\right.\nonumber \\ & & \left. 
+\frac{m_\pi}{\sqrt{4M_0^2-m_\pi^2}}\left(
14-8\frac{m_\pi^2}{M_0^2}+\frac{m_\pi^4}{M_0^4}
\right)\arccos\frac{m_\pi}{2M_0}
\right]
\\ &&
+{ \Delta a_{20}^{\rm v}}
\frac{g_A^2m_\pi^2}{48\pi^2F_\pi^2}\left[
2\frac{m_\pi^2}{M_0^2}+\frac{m_\pi^2}{M_0^2}\left(
6-\frac{m_\pi^2}{M_0^2}\right)\ln\frac{m_\pi^2}{M_0^2}
+2m_\pi\frac{(4M_0^2-m_\pi^2)^{3/2}}{M_0^4}\arccos\frac{m_\pi}{2M_0}
\right] \nonumber
\end{eqnarray}
for the isovector channel, where $M_0$ is the nucleon mass in the chiral limit.
We perform 2-parameter fits with $a_{20}^{\rm s}$, $c_9$ for the isoscalar
and $a_{20}^{\rm v}$, $c_8$ for the isovector channel 
at a scale of $\lambda=1$GeV and fixed the other values
following ref.\cite{DGH}.
A strong $m_\pi$ dependence is observed especially for the isovector channel,
but the extrapolated values in both channels overshoot beyond the 
phenomenological values estimated using the CTEQ6 parton distribution 
functions.
The chiral extrapolation gives 
$A_{20}^{\rm u+d}\equiv\langle x\rangle^{\rm u+d}
= 0.572\pm 0.012$ for the isoscalar channel and
$A_{20}^{\rm u-d}\equiv\langle x\rangle^{\rm u-d}
= 0.198\pm 0.008$ for the isovector channel at the physical pion mass. 
See ref.\cite{DP} for discretization effects of these form factors.

\begin{figure}[t!]
\centering
{\includegraphics[width=.35\textwidth,angle=270]{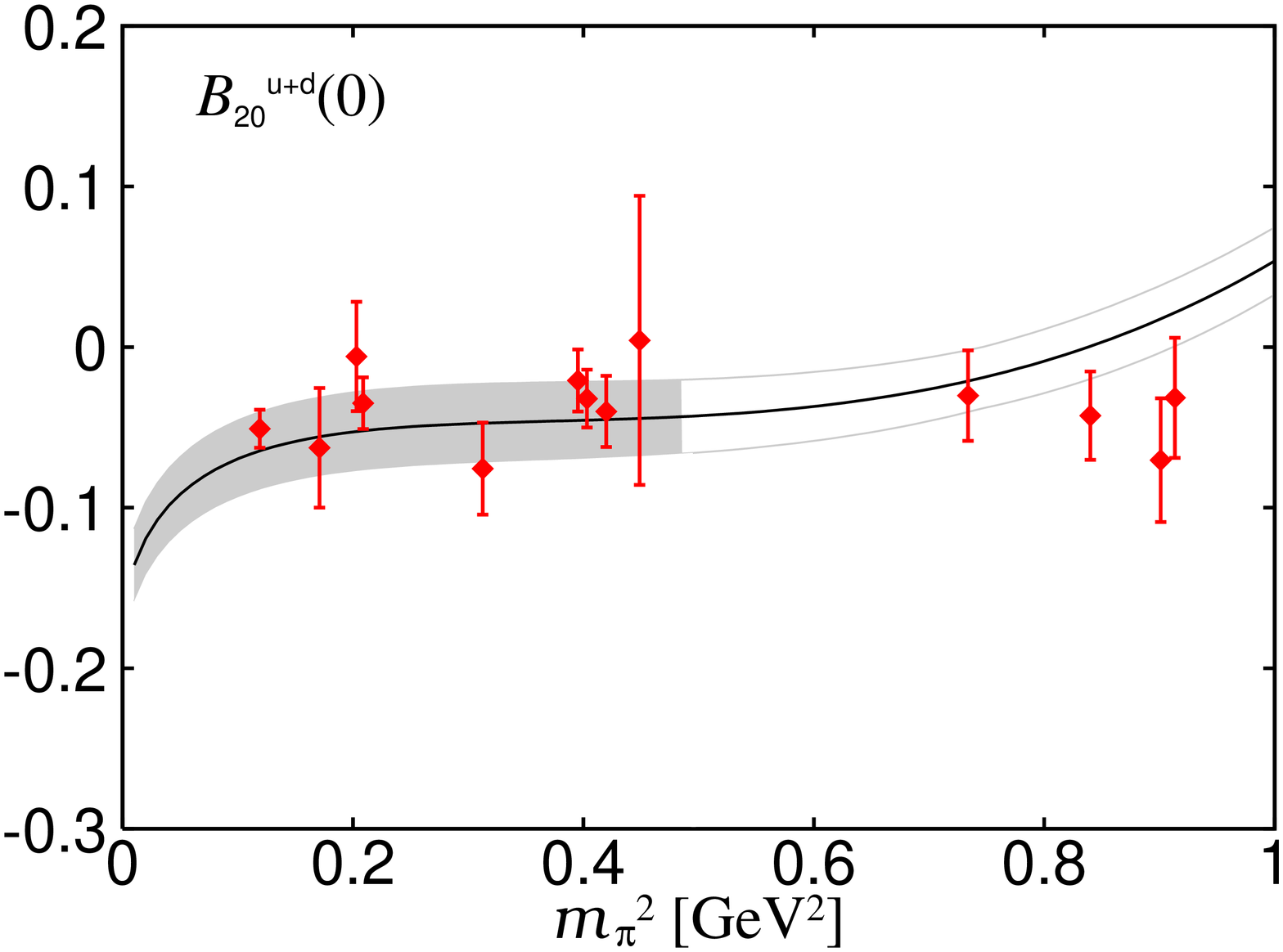} \hspace*{1.6em}
\includegraphics[width=.35\textwidth,angle=270]{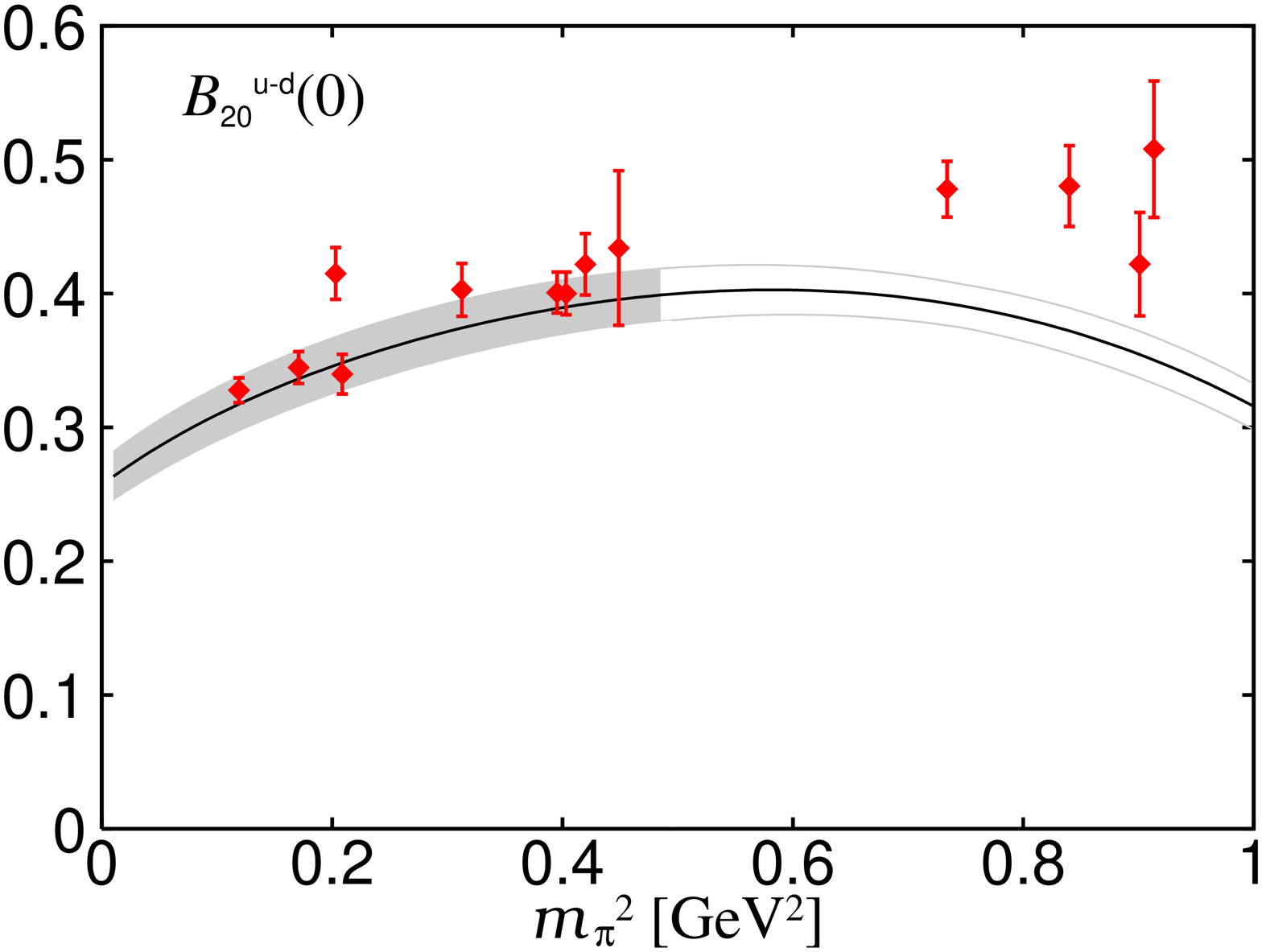}}
  \caption{ The forward values of $B_{20}$ extrapolated
by $\chi$PT in the isoscalar (left) and isovector (right) channel 
with $\chi$PT fits.}
  \label{fig:B20}
\end{figure}
 In contrast to $A_{20}$, the forward values of $B_{20}$ cannot be calculated
directly from the lattice simulation since the kinematic pre-factor
for $B_{20}$ vanishes at zero momentum transfer. Again we use the
expressions of covariant baryon $\chi$PT \cite{DGH},
\begin{eqnarray}
B_{2,0}^{\rm u\pm d}({t}) &=&  ({ b_{20}^{\rm s,v}
}+{ \hat{\delta}_B^{\rm s,v}}\, m_\pi^2 + 
{ \hat{\delta}_{Bt}^{\rm s,v}}\, { t})\,\frac{M_{\rm N}(m_{\pi})}{M_0}\mp
a_{20}^{\rm s,v} \frac{(2\pm 1)g_A^2M_0^2}{48\pi^2F_{\pi}^2} G({ t}), \\
 G({ t})&=&
\int_{-\frac{1}{2}}^{\frac{1}{2}}\!
        \frac{du}{\tilde{M}^8}\Bigg[\left(M_0^2-\tilde{M}^2\right)\tilde{M}^6+
9m_{\pi}^2M_0^2\tilde{M}^4
 -6m_{\pi}^4M_0^2\tilde{M}^2+6m_{\pi}^2M_0^2
\left(m_{\pi}^4-3m_{\pi}^2\tilde{M}^2+\tilde{M}^4\right)\ln{\frac{m_{\pi}}{\tilde{M}}}
\nonumber \\  && \left. \hspace*{4em}
-\frac{6m_{\pi}^3M_0^2}{\sqrt{4\tilde{M}^2-m_{\pi}^2}}\bigg(
        m_{\pi}^4-5m_{\pi}^2\tilde{M}^2+5\tilde{M}^4\bigg)
\arccos{\frac{m_{\pi}}{2\tilde{M}}}\Bigg]
\right|_{{\tilde{M}^2}=M_0^2+\left(u^2-\frac{1}{4}\right) t}, \nonumber
\end{eqnarray}
to fit the lattice results as a function of $t$ and $m_\pi$.
Free fit parameters are  $b_{20}^{\rm s,v}$,
${\hat{\delta}_B^{\rm s,v}}$ and ${\hat{\delta}_{Bt}^{\rm s,v}}$
for the isoscalar and isovector channel respectively, and 
the parameters of $a_{20}^{\rm s,v}$ are fixed by the fitting 
of $A_{20}^{\rm u\pm d}$.
The forward values of $B_{20}$ extracted from this fit
with fixed $m_\pi$ are shown in Fig.\ref{fig:B20}, where
the solid lines represent
the section of fitting surfaces at $t=0$.

 Since the forward value of $B_{20}^{\rm u+d}$ is equivalent to
the difference of $2J^q-\langle x\rangle^q$, the small values of
the lattice results indicate
 the cancellation between total angular momentum
and momentum fraction of quarks. However, the $\chi$PT fit suggests
a sizeable bending through the chiral extrapolation, which makes
a sharp contrast to the chiral quark soliton model \cite{CQSM}.
We obtain $B_{20}^{\rm u+d}(0)=-0.120\pm 0.023$ at the physical point
for the isoscalar channel and $B_{20}^{\rm u-d}(0)=0.269\pm 0.020$ 
for the isovector channel.

\begin{figure}[t!]
\centering
{\includegraphics[width=.35\textwidth,angle=270]{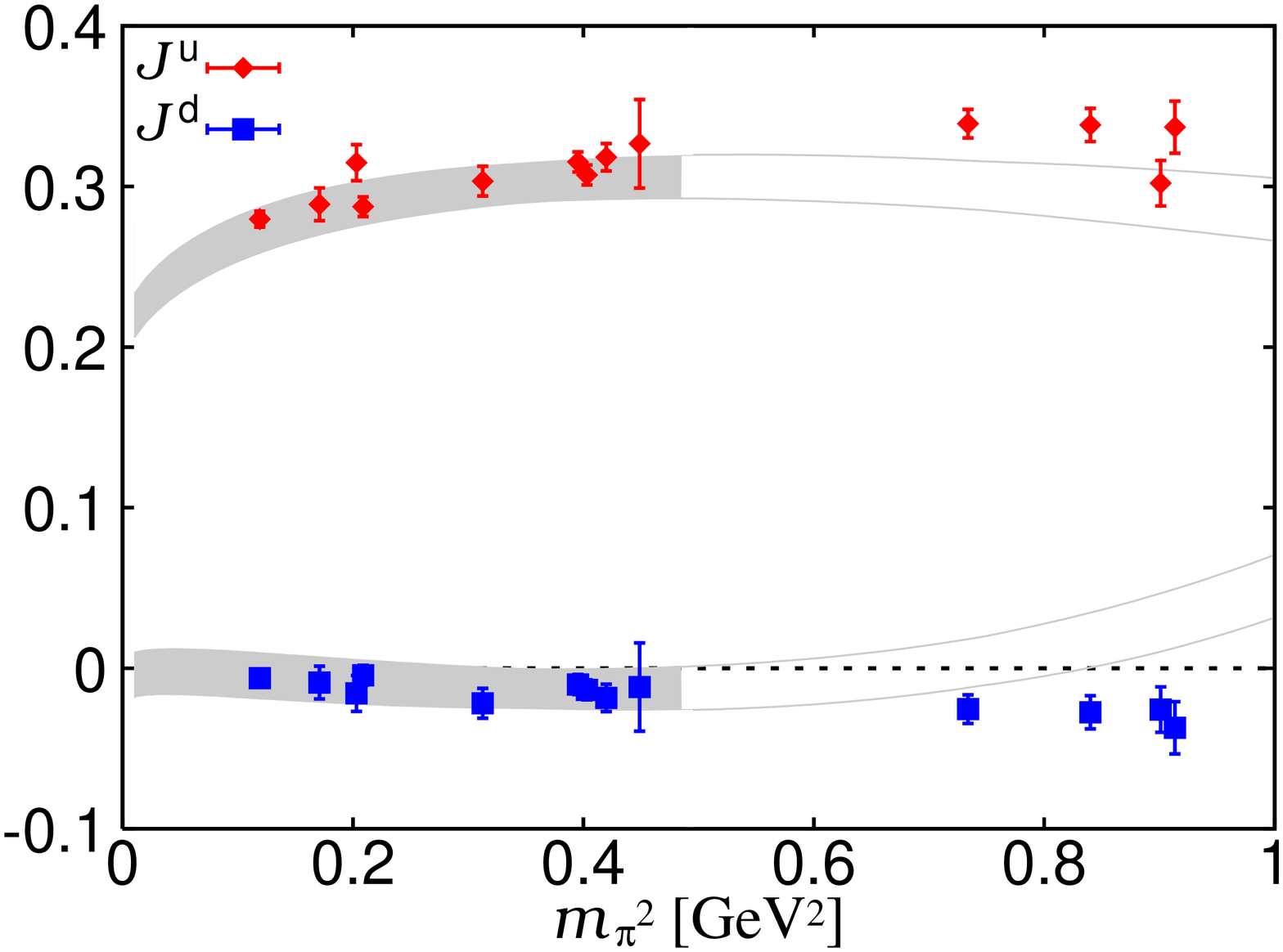} \hspace*{1.6em}
\includegraphics[width=.35\textwidth,angle=270]{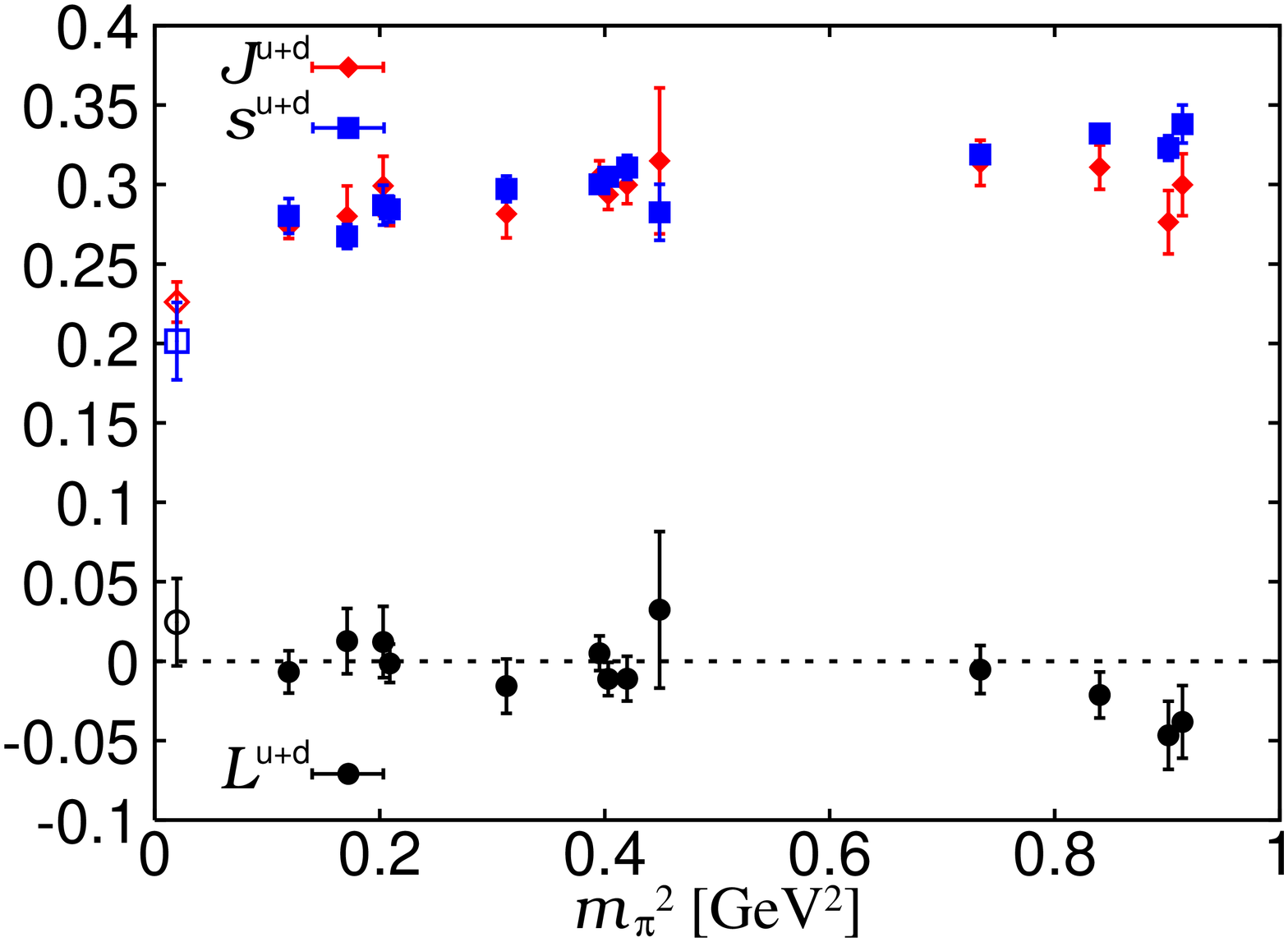}}
  \caption{ Total angular momentum of quark in the nucleon with $\chi$PT fit
(left) and spin, orbital angular momentum of quarks (right). The open
symbols represent the extrapolated values to the physical pion mass.}
  \label{fig:Jq}
\end{figure}

 Combining all these data, we can estimate the angular momentum of quarks
in the nucleon. The pion mass dependence of
the total angular momentum $J^q$ for u and d quark is shown in Fig.\ref{fig:Jq}.
The strong $m_\pi$ dependences of $A_{20}$ in the isovector channel and
$B_{20}$ in the isoscalar channel are enhanced for the u quark and gives
a significant suppression of $J^{\rm u}$ near the physical point, while these
dependences cancel each other for the d quark and so keep the value of
$J^{\rm d}$ small.
 The extrapolation to the physical point gives $J^{\rm u}= 0.230\pm 0.008$
and $J^{\rm d}=-0.004\pm 0.008$.

 The quark spin $s^q$, the total and orbital angular momentum $L^q=J^q-s^q$ 
are shown in the right panel of Fig.\ref{fig:Jq}.
We obtain the values at physical point by the chiral extrapolation 
as $J^{\rm u+d}=0.226\pm 0.013$, $s^{\rm u+d}=0.201\pm 0.024$ and
$L^{\rm u+d}=0.025\pm 0.027$.
 The results show that the total angular momentum of 
quark is comparable with the quark spin and hence the orbital angular
momentum is consistent with zero, which agrees with the result of
ref.\cite{PH}.

\section{Conclusions}

 We have carried out lattice simulations to calculate
the first moments of GPDs, which play an important role for 
the proton spin, quark transverse density and deeply virtual
Compton scattering.
 The lattice results of the generalized form factor $A_{20}$
are fitted to the dipole form for small $-t$ region and the dipole
mass turns out to be comparable with the observed tensor meson mass.
Since this empirical fit has no solid justification from a theoretical 
point of view,
 we employ baryon $\chi$PT 
to take the forward limit of $B_{20}$ and to
chirally extrapolate the form factors to the physical pion mass.
 The resulting values indicate that the total angular
momentum of quarks in the nucleon is of the same size
as the quark spin contribution, while the orbital
angular momentum is consistent with zero.

Further analyses are needed to estimate the finite size effects \cite{DL},
contributions from disconnected diagrams and so on.
Results with lighter pion masses will be forthcoming. 

\acknowledgments{
The numerical calculations have been performed on the Hitachi SR8000
at LRZ (Munich), the BlueGene/L and the Cray T3E at EPCC (Edinburgh),
the BlueGene/Ls at NIC/FZJ (J\"ulich) and KEK (by the Kanazawa group
as part of DIK research program) and on the APEmille and apeNEXT at
NIC/DESY (Zeuthen).
This work was supported in part by the DFG and by   
the EU Integrated
Infrastructure Initiative Hadron Physics (I3HP) under contract 
number RII3-CT-2004-506078.
}

\end{document}